\documentclass[a4paper,12pt]{article}
\usepackage[T2A]{fontenc}
\usepackage[cp1251]{inputenc}  
\usepackage[russian, english]{babel}
\usepackage {euscript}
\usepackage{epsfig}
\usepackage{amsmath}
\usepackage{amssymb}
\usepackage{bm}
\usepackage{indentfirst}

\newcommand{\beq}{\begin{equation}}
\newcommand{\eeq}{\end{equation}}
\newcommand{\beqn}{\begin{eqnarray}}
\newcommand{\eeqn}{\end{eqnarray}}

\begin{document}

\begin{center}
{\bf \large Electromagnetic production of positron and electron\\
in collisions of heavy nuclei}
\end{center}

\begin{center}
I.B. Khriplovich\footnote{khios231@mail.ru}
\end{center}
\begin{center}
St.Petersburg State University, 198504 St.Petersburg, Russia
\end{center}

\bigskip

\begin{abstract}

We consider the electromagnetic production of positron and
electron in collisions of slow heavy nuclei. This process is
dominated by emission of positron, with the electron captured by
nucleus.

\end{abstract}

\vspace{8mm}

Positron production in collisions of heavy nuclei, as well as the
production $e^+ e^-$ pairs, were addressed in numerous papers
(see, for instance, [1,2]). Here we consider this problem with
simple qualitative arguments.

\begin{figure}[h!]
\centering
\caption{
Production of pair $e^+ e^-$.
}
\includegraphics[width = 0.5\textwidth]
{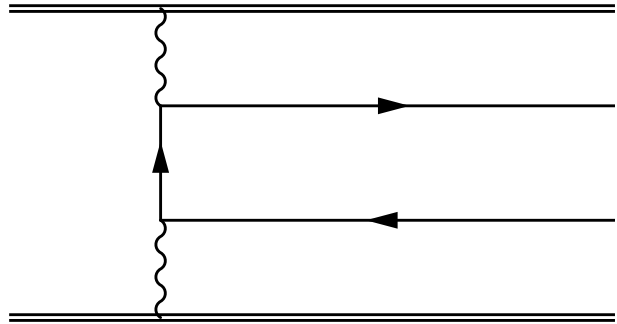} 
\label{fig:1a}
\end{figure}

We start with production of the pair $e^+ e^-$, as presented in
Fig.~\ref{fig:1a} 
Double lines therein describe the propagation of heavy
nuclei, single and wavy lines refer to $e^\pm$ and virtual
photons, respectively. The electromagnetic vertices on these
double lines reduce to $(p^\prime + p)/2M \approx v$. We neglect 
the difference between $p^\prime$ and $p$. For the velocity of nuclei we assume $v \simeq 0.1$. Thus, taking into
account both nuclear lines in Fig. 1, we conclude that our
amplitude is proportional to $v^2$. As to the discussed
cross-section, it is obviously proportional to
\beq
v^4/v = v^3;
\eeq
$v$ in this denominator originates from the particle flux; here
and below we put $c=1$.

\begin{figure}[h!]
\centering
\caption{
Capture of electron.
}
\includegraphics[width = 0.5\textwidth]
{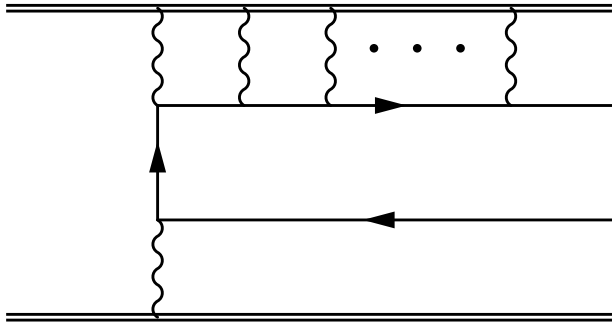} 
\label{fig:2}
\end{figure}

We address now the channel where the produced electron is
captured by one of the nuclei, and the produced positron goes to
infinity, as presented in Fig.~\ref{fig:2}. 
The captured electron here has
no direct relation at all to the velocity discussed. Indeed, the
attraction of this electron to nucleus is strong, and there is no
reason to expect any extra $v$-dependence in the corresponding
vertex. Therefore, the amplitude here is proportional to $v$, 
with the corresponding cross-section proportional to
\beq
v^2/v = v.
\eeq

Thus, the positron production cross-section dominates the effect,
since
\beq
v/v^3 \simeq 10^2.
\eeq
In other words, the cross-section of the positron production, with
the capture of the produced electron, is the dominating one, it is about hundred times more
than that of $e^+ e^-$ production. The attraction of the produced
electron to nucleus is strong, and there is no reason to expect
any extra $v$-dependence in the corresponding vertex.

The electromagnetic cross-section of $e^+ e^-$ production was
estimated in [3] as
\beq
\sigma_{\pm} \sim 10^{-26} - 10^{-25}\, \rm{cm}^2.
\eeq
Then, the reasonable estimate for the cross-section of 
positron production is
\beq
\sigma_+ \sim 10^{-24} - 10^{-23}\, \rm{cm}^2.
\eeq

\subsection*{Acknowledgements}

I am grateful to V.M. Shabaev for useful discussions.

\renewcommand{\bibname}

{\normalsize References}
\begin{thebibliography}\\

\bibitem{mu}
U. Mueller, T.D. Reus, J. Reinhardt, B. Mueller, W. Greiner, G.
Soff,\\ Phys. Rev. A 37 1449 (1988).

\bibitem{ma}
I.A. Maltsev, V.M. Shabaev, I.I. Tupitsin, A.I. Bondarev, Y.S.
Kozhedub, G. Plunien, Th. Stolker, Phys. Rev. A 91  032708 (2015).

\bibitem{kh}
I.B. Khriplovich, JETP Letters 100 552 (2014).

\end{thebibliography}
\end{document}